\documentclass[11pt,a4paper]{article}
\pdfoutput=1
\usepackage{amssymb,latexsym,amsmath}
\usepackage[dvipdfmx]{graphicx}
\usepackage{enumerate}
\usepackage{mysty2e_eng}
\usepackage{color}
\usepackage{txfonts}

\definecolor{brown}{rgb}{0.8,0.4,0}
\definecolor{purple}{rgb}{0.49,0.18,0.56}

\begin{document}

\def\varliminf{\mathop{\underline{\rule[-0.02em]{0em}{0.2em}
 \hbox{\rm lim}}}}
\def\varlimsup{\mathop{\overline{\hbox{\rm lim}}}}
\def\esssup{\mathop{\hbox{\rm ess\,sup}}}


\newcounter{thebrenum}
\newenvironment{brenum}%
{\begin{list}{(\arabic{thebrenum}) }%
{\usecounter{thebrenum}
\setcounter{thebrenum}{0}
\setlength{\labelsep}{0ex}
\setlength{\labelwidth}{4.1ex}
\setlength{\leftmargin}{\labelwidth}
\setlength{\itemsep}{0ex}
\setlength{\parsep}{0.5ex}
}}%
{\end{list}}


\newcounter{thebrenumalph}
\newenvironment{brenumalph}%
{\begin{list}{(\alph{thebrenumalph}) }%
{\usecounter{thebrenumalph}
\setcounter{thebrenumalph}{0}
\setlength{\labelsep}{0ex}
\setlength{\labelwidth}{4.1ex}
\setlength{\leftmargin}{\labelwidth}
\setlength{\itemsep}{0ex}
\setlength{\parsep}{0.5ex}
}}%
{\end{list}}


\newcounter{thebrenumroman}
\newenvironment{brenumroman}%
{\begin{list}{(\roman{thebrenumroman}) }%
{\usecounter{thebrenumroman}
\setcounter{thebrenumroman}{0}
\setlength{\labelsep}{0ex}
\setlength{\labelwidth}{3.3ex}
\setlength{\leftmargin}{\labelwidth}
\setlength{\itemsep}{0ex}
\setlength{\parsep}{0.7ex}
}}%
{\end{list}}


\newcommand{\bda}{\boldsymbol{a}}
\newcommand{\bdb}{\boldsymbol{b}}
\newcommand{\bdc}{\boldsymbol{c}}
\newcommand{\bdd}{\boldsymbol{d}}
\newcommand{\bde}{\boldsymbol{e}}
\newcommand{\bdf}{\boldsymbol{f}}
\newcommand{\bdg}{\boldsymbol{g}}
\newcommand{\bdh}{\boldsymbol{h}}
\newcommand{\bdi}{\boldsymbol{i}}
\newcommand{\bdj}{\boldsymbol{j}}
\newcommand{\bdk}{\boldsymbol{k}}
\newcommand{\bdl}{\boldsymbol{l}}
\newcommand{\bdm}{\boldsymbol{m}}
\newcommand{\bdn}{\boldsymbol{n}}
\newcommand{\bdo}{\boldsymbol{o}}
\newcommand{\bdp}{\boldsymbol{p}}
\newcommand{\bdq}{\boldsymbol{q}}
\newcommand{\bdr}{\boldsymbol{r}}
\newcommand{\bds}{\boldsymbol{s}}
\newcommand{\bdt}{\boldsymbol{t}}
\newcommand{\bdu}{\boldsymbol{u}}
\newcommand{\bdv}{\boldsymbol{v}}
\newcommand{\bdw}{\boldsymbol{w}}
\newcommand{\bdx}{\boldsymbol{x}}
\newcommand{\bdy}{\boldsymbol{y}}
\newcommand{\bdz}{\boldsymbol{z}}
\newcommand{\bdH}{\boldsymbol{H}}
\newcommand{\bdN}{\boldsymbol{N}}
\newcommand{\bdR}{\boldsymbol{R}}
\newcommand{\bdS}{\boldsymbol{S}}
\newcommand{\bdL}{\boldsymbol{L}}

\newcommand{\bdzero}{\boldsymbol{0}}
\newcommand{\bdone}{\boldsymbol{1}}

\newcommand{\bdalpha}{\boldsymbol{\alpha}}
\newcommand{\bdbeta}{\boldsymbol{\beta}}
\newcommand{\bdtheta}{\boldsymbol{\theta}}
\newcommand{\bdlambda}{\boldsymbol{\lambda}}
\newcommand{\bdxi}{\boldsymbol{\xi}}
\newcommand{\bdeta}{\boldsymbol{\eta}}
\newcommand{\bdzeta}{\boldsymbol{\zeta}}
\newcommand{\bdomega}{\boldsymbol{\omega}}

\newcommand{\rma}{{\mathrm a}}
\newcommand{\rmb}{{\mathrm b}}
\newcommand{\rmc}{{\mathrm c}}
\newcommand{\rmd}{{\mathrm d}}
\newcommand{\rme}{{\mathrm e}}
\newcommand{\rmf}{{\mathrm f}}
\newcommand{\rmg}{{\mathrm g}}
\newcommand{\rmh}{{\mathrm h}}
\newcommand{\rmi}{{\mathrm i}}
\newcommand{\rmj}{{\mathrm j}}
\newcommand{\rmk}{{\mathrm k}}
\newcommand{\rml}{{\mathrm l}}
\newcommand{\rmm}{{\mathrm m}}
\newcommand{\rmn}{{\mathrm n}}
\newcommand{\rmo}{{\mathrm o}}
\newcommand{\rmp}{{\mathrm p}}
\newcommand{\rmq}{{\mathrm q}}
\newcommand{\rmr}{{\mathrm r}}
\newcommand{\rms}{{\mathrm s}}
\newcommand{\rmt}{{\mathrm t}}
\newcommand{\rmu}{{\mathrm u}}
\newcommand{\rmv}{{\mathrm v}}
\newcommand{\rmw}{{\mathrm w}}
\newcommand{\rmx}{{\mathrm x}}
\newcommand{\rmy}{{\mathrm y}}
\newcommand{\rmz}{{\mathrm z}}

\newcommand{\rmA}{{\mathrm A}}
\newcommand{\rmE}{{\mathrm E}}
\newcommand{\rmF}{{\mathrm F}}
\newcommand{\rmG}{{\mathrm G}}
\newcommand{\rmH}{{\mathrm H}}
\newcommand{\rmL}{{\mathrm L}}
\newcommand{\rmM}{{\mathrm M}}
\newcommand{\rmN}{{\mathrm N}}
\newcommand{\rmO}{{\mathrm O}}
\newcommand{\rmR}{{\mathrm R}}
\newcommand{\rmT}{{\mathrm T}}
\newcommand{\rmU}{{\mathrm U}}

\newcommand{\vGamma}{\varGamma}
\newcommand{\vDelta}{\varDelta}
\newcommand{\vTheta}{\varTheta}
\newcommand{\vLambda}{\varLambda}
\newcommand{\vXi}{\varXi}
\newcommand{\vPi}{\varPi}
\newcommand{\vSigma}{\varSigma}
\newcommand{\vUpsilon}{\varUpsilon}
\newcommand{\vPhi}{\varPhi}
\newcommand{\vPsi}{\varPsi}
\newcommand{\vOmega}{\varOmega}

\newcommand{\bdvDelta}{{\boldsymbol{\varDelta}}}

\allowdisplaybreaks
\newcommand{\setunit}[1]{\setlength{\unitlength}{#1}}
\renewcommand{\hat}[1]{\widehat{#1}}
\renewcommand{\tilde}[1]{\widetilde{#1}}

\newcommand{\qed}{\hfill$\square$}
\newcommand{\fin}{\hfill$\Diamond$}
\newcommand{\etal}{{\em et al.}}
\newcommand{\ie}{{\em i.e.}}
\newcommand{\eg}{{\em e.g.}}

\newcommand{\Zset}{\mathbb Z}
\newcommand{\Rset}{\mathbb R}
\newcommand{\Cset}{\mathbb C}
\newcommand{\Msng}{\overline{\sigma}}
\newcommand{\msng}{\underline{\sigma}}
\newcommand{\Meig}{\overline{\lambda}}
\newcommand{\meig}{\underline{\lambda}}
\newcommand{\ol}[1]{\overline{#1}}
\newcommand{\ul}[1]{\underline{#1}}

\newtheorem{thm}{Theorem}
\newtheorem{prop}[thm]{Proposition}
\newtheorem{lem}[thm]{Lemma}
\newtheorem{cor}[thm]{Corollary}
\newtheorem{problem}[thm]{Problem}
\newenvironment{prob}{\begin{problem}\upshape}{\end{problem}}
\newtheorem{algorithm}{Algorithm}
\newenvironment{algo}{\begin{algorithm}\upshape}{\end{algorithm}}
\newtheorem{example}{Example}
\newenvironment{exmpl}{\begin{example}\upshape}{\end{example}}
\newtheorem{remark}{Remark}
\newenvironment{rem}{\begin{remark}\upshape}{\end{remark}}
\newtheorem{assumption}{Assumption}
\newenvironment{assum}{\begin{assumption}\upshape}{\end{assumption}}

\title{\vspace*{-1.2cm}\bf\Large
Optimal Sampled-Data Control of a Nonlinear System$^*$%
\footnotetext{$\mbox{}^*$%
Released on December 29, 2021.
Preprint submitted to {\em Automatica}.
This work is supported by the Nanzan University Pache Research Subsidy
I-A-2 for the academic year 2021.}%
\footnotetext{$\mbox{}^\dagger$%
Department of Mechanical Engineering and System Control,
Nanzan University,
Yamazatocho 18, Showa-ku, Nagoya 466-8673, Japan;
email: oishi@nanzan-u.ac.jp, noboru.sakamoto@nanzan-u.ac.jp}%
\vspace*{-0.2cm}}
\author{\large Yasuaki Oishi$^\dagger$ and Noboru Sakamoto$^\dagger$}
\date{\normalsize\vspace*{-0.9cm}}

\abstract{%
\small\setlength{\baselineskip}{14.5pt}\noindent
Optimal sampled-data control of a nonlinear system is considered
with the stable-manifold approach and extensive use of numerical techniques.
The idea is to notice the Hamiltonian system associated with
the considered optimal control problem and
to compute trajectories on its stable manifold.
Since the control input accompanied with those trajectories is
proved to be optimal,
the optimal control law can be obtained through interpolation.
The stable-manifold approach was originally proposed for
continuous-time optimal control and here it is adapted
for sampled-data control based on the works of Navasca.
In the case of sampled-data control,
the approach requires the state transition of the controlled plant
during one sampling period together with its derivatives
with respect to the state and the input.
Their computation is achieved by numerical techniques.
Moreover, a shooting method is proposed for systematic generation
of the trajectories and
extension is considered for the intersample behavior to be
taken into account.
The proposed method is applied to tracking control of a wheeled mobile robot.
It works successfully with a rather long sampling period.
\\[0.2cm]%
\vspace*{-0.1cm}}

\maketitle
\thispagestyle{empty}

\section{Introduction}
\label{sec:intro}
A sampled-data control system is a system where
a continuous-time plant is controlled by a discrete-time controller
together with sampling and hold devices.
Nowadays, most of control systems are sampled-data control systems
because a controller is usually implemented with a digital technique.
Sampled-data control of a linear plant has been thoroughly investigated and
a methodology for the optimal control is established
according to various performance indices
possibly with intersample behavior taken into account \cite{ChF95}.

In contrast, sampled-data control of a nonlinear system
is far from matured and most of the existing control methods
are based on approximation.
One approach is to design a continuous-time controller first for
a given continuous-time plant and then to approximate it
by a discrete-time controller \cite{NeG05,NeG06}.
Another approach is to discretize a given continuous-time plant
by the Euler approximation or something similar and then
to design a discrete-time controller \cite{NTK99,LNA06,NeT06,KaA14}.
In both approaches, the sampling period has to be sufficiently small
for the approximation to be valid.
In \cite{BoT15,BoT17}, dynamic programming of a sampled-data control system
is considered in a general framework without approximation.
The authors do not know however an existing method that gives
an optimal sampled-data control law in a practical state-feedback form
without resorting to approximation.
One of the difficulties should be that
the state transition during one sampling period is hard to obtain
analytically for a nonlinear plant.

In this paper,
we consider optimal sampled-data control of a nonlinear plant
with the stable-manifold approach and extensive use
of numerical techniques.
The stable-manifold approach is originally developed for continuous-time control
and has been applied to various practical nonlinear control problems
\cite{YaS98,SaS08,Sak13}.
There, the Hamiltonian system is considered for a given optimal control problem
and its trajectories on the stable manifold are computed.
Since a trajectory on the stable manifold gives a correspondence
between a state and the optimal input,
one can have an optimal control law in a state-feedback form
through interpolation.
In order to apply this approach to sampled-data control,
one can utilize the results of Navasca \cite{Nav02,Nav03},
where the Hamiltonian system is considered for discrete-time control
and its connection to the optimal control is discussed.
Computation of the optimal sampled-data control law is however
not straightforward.
Since the Hamiltonian system in \cite{Nav02,Nav03} is implicit
both in forward time and backward time,
a numerical technique is necessary to compute its trajectory.
The technique further requires the state transition of the controlled
plant and its derivative with respect to the initial state
and the given control input.
With a numerical technique, again, they can be computed and
this is the basis of the proposed method.
Although the control performance is evaluated only at the sampling instants first,
extension is possible to take care of the intersample performance.
There, a numerical technique again plays a central role.

The rest of the paper is structured as follows.
Section~\ref{sec:prob} presents the optimal control problem to be considered.
After some preliminaries in Section~\ref{sec:prel},
the Hamiltonian system is presented in Section~\ref{sec:Hamilton}.
Section~\ref{sec:method} gives a method to solve the posed problem
and Section~\ref{sec:shooting} provides a shooting method
for efficient generation of a trajectory.
Optimal control taking care of intersample performance
is considered in Section~\ref{sec:inter}.
After a numerical example is given in Section~\ref{sec:example},
the paper is concluded in Section~\ref{sec:concl}. 

The following notation is used.
The symbol $\Rset$ stands for the set of real numbers and
$\Rset^n$ for the set of $n$-dimensional real column vectors. 
The transpose of a vector and a matrix is denoted by $\mbox{}^\rmT$.
The derivative of a scalar-valued function $H$ with respect to
a vector variable $u=(u_1\ \ \cdots\ \ u_m)^\rmT$ is
an $m$-dimensional row vector $\partial H/\partial u$
whose $i$th component is $\partial H/\partial u_i$.
The derivative of an $n$-dimensional vector-valued function
$\phi=(\phi_1\ \ \cdots\ \ \phi_n)^\rmT$
with respect to $u=(u_1\ \ \cdots\ \ u_m)^\rmT$ is
an $n\times m$ matrix $\partial\phi/\partial u$
whose $(i,j)$-component is $\partial\phi_i/\partial u_j$.
Moreover, the second-order derivative of a scalar-valued function $H$
with respect to $u=(u_1\ \ \cdots\ \ u_m)^\rmT$ and $x=(x_1\ \ \cdots\ \ x_n)^\rmT$
is an $m\times n$ matrix $\partial^2 H/\partial u\partial x$
whose $(i,j)$-component is
$\partial^2 H/\partial u_i\partial x_j$.

\section{Considered problem}
\label{sec:prob}
\begin{figure}[t]
\begin{center}
\setlength{\unitlength}{1cm}
\begin{picture}(0,3.5)(0,0.5)
\put(0,0){\makebox(0,0)[b]{\includegraphics[height=3.5cm]{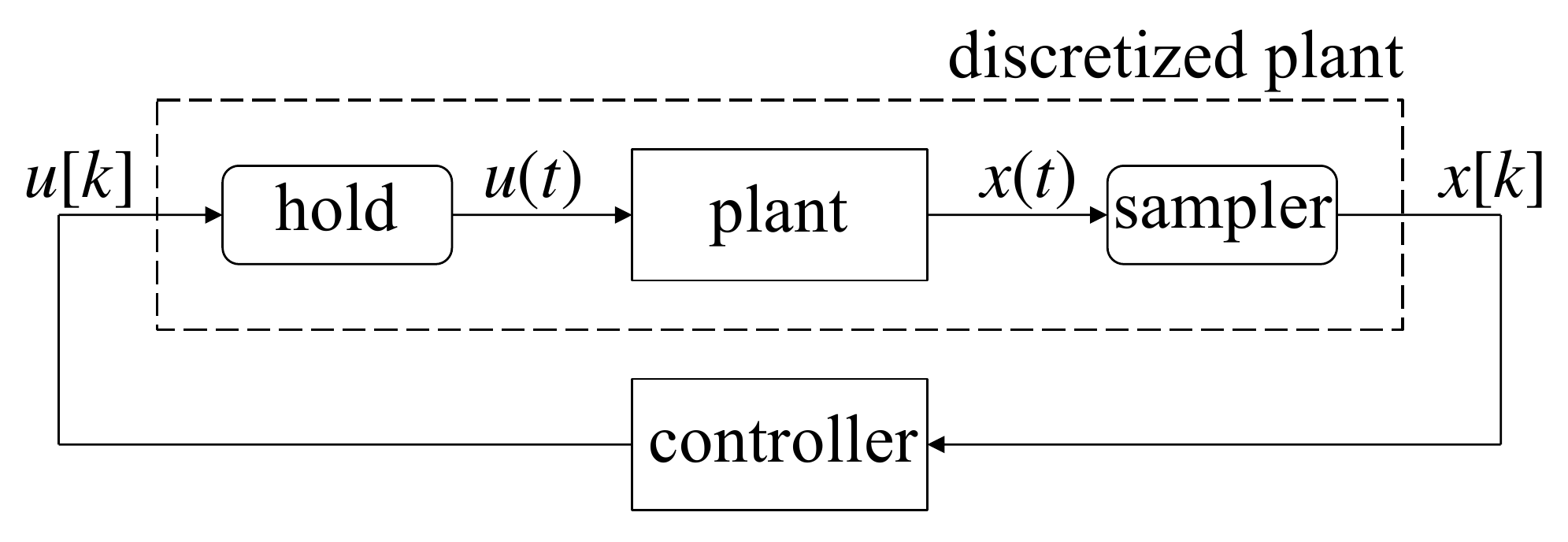}}}
\end{picture}
\end{center}
\caption{The sampled-data control system to be considered.
The portion enclosed by the dashed rectangle can be
regarded as a discrete-time system and is called
the discretized plant.}
\label{fig:system}
\end{figure}

A plant to be controlled is a continuous-time system
\begin{equation}
\dot{x}(t)=f(x(t),u(t))
\label{eq:original}
\end{equation}
with the state $x(t)\in\Rset^n$ and the input $u(t)\in\Rset^m$.
Here, the function $f(x,u)$ is assumed to be sufficiently smooth
and to satisfy $f(0,0)=0$.
We consider to control this plant with a discrete-time controller
that produces a discrete-time input $u[k]$ from a sampled state $x[k]$.
See Figure~\ref{fig:system}.
Here, $x[k]$ is the state $x(t)$ sampled at $t=kh$, $k=0, 1, \ldots$,
for some sampling period $h>0$
and $u[k]$ is converted to the continuous-time input $u(t)$
by $u(t)=u[k]$ for $kh\leq t<(k+1)h$ and $k=0, 1, \ldots$.

Let us consider the portion enclosed by the dashed rectangle in the figure,
which is regarded as a discrete-time system and is called the discretized plant.
To have its expression, consider the initial-value problem
\begin{equation}
\dot{x}(t)=f(x(t),u_0),\quad x(0)=x_0.
\label{eq:initial}
\end{equation}
We assume that this initial-value problem
has a unique solution $\phi(t; x_0, u_0)$, $0\leq t\leq h$,
for any $x_0\in X$ and any $u_0\in U$,
where $X\subset\Rset^n$ and $U\subset\Rset^m$ are
some bounded open sets containing the origin and
$U$ is also convex.
It is known \cite[Chapter 6]{Sid13}
that $\phi(t;x_0,u_0)$ is smooth in $(t,x_0,u_0)$.
Now, the discrete-time system
\begin{equation}
x[k+1]=\phi(h;x[k],u[k])
\label{eq:discretized}
\end{equation}
is the desired discretization of the plant \eqref{eq:original}.

Our problem to be considered is the following.
\bigskip

\noindent
{\bf Problem.}
\label{prob:optimal}
Let $Q$ be some positive semidefinite matrix and $R$ be
some positive definite matrix.
Obtain the optimal input $u[k]\in U$ for each $k=0, 1, \ldots$
that minimizes the objective function
\begin{equation}
\sum_{k=0}^\infty
\Bigl(hx[k]^\rmT Qx[k]+hu[k]^\rmT Ru[k]\Bigr).
\label{eq:objective}
\end{equation}
\fin

\begin{rem}
\label{rem:consistency}
To the objective function above, the sampling period $h$ is multiplied.
It is for consistency with the objective function
in Section~\ref{sec:inter},
where the intersample behavior of the system is considered.
\fin
\end{rem}

\section{Preliminaries}
\label{sec:prel}
Given a nonlinear function $f(x,u)$,
it is hard to obtain an analytic form of $\phi(t;x,u)$
in the initial-value problem \eqref{eq:initial}.
For a specific value of $(t,x,u)$, however,
the value of $\phi(t;x,u)$ can be computed by a numerical technique.
The same is true for the derivatives of $\phi(t;x,u)$.
The details are explained in this section as a preliminary for
the rest of the paper.

Suppose that some $(x,u)\in X\times U$ is given.
By definition, the function $\phi(t;x,u)$ satisfies
\begin{equation}
\frac{\rmd}{\rmd t}\phi(t;x,u)=f(\phi(t;x,u),u),\quad
\phi(0;x,u)=x.
\label{eq:eq4phi}
\end{equation}
We are to apply a numerical technique
to compute $\phi(t;x,u)$ for various $t$ between $0$ and $h$.
One example of such technique is 
the classical fourth-order Runge--Kutta method \cite[Section 12.5]{SuM03}.
There, with $M$ being some positive integer and
$\hat{\phi}_0:=x$, we repeat the following procedure for $j=0, 1, \ldots, M-1$:
\begin{align*}
g_1&:=f(\hat{\phi}_j,u),\\
g_2&:=f(\hat{\phi}_j+\frac{h}{2M}g_1,u),\\
g_3&:=f(\hat{\phi}_j+\frac{h}{2M}g_2,u),\\
g_4&:=f(\hat{\phi}_j+\frac{h}{M}g_3,u),\\
\hat{\phi}_{j+1}&:=\hat{\phi}_j+\frac{h}{6M}(g_1+2g_2+2g_3+g_4).
\end{align*}
Then, $\hat{\phi}_j$ gives the value of $\phi(jh/M;x,u)$
for $j=1, 2, \ldots, M$.
For the moment, we need only the final value $\phi(h;x,u)$
that appears in the expression \eqref{eq:discretized}
of the discretized plant.
Later in Section~\ref{sec:inter}, however, we also use the
intermediate values $\phi(jh/M;x,u)$ to assess
the intersample behavior of the system.

For computation of the derivative of $\phi(t;x,u)$
with respect to $x$ and $u$,
differentiate both sides of \eqref{eq:eq4phi} as
\begin{alignat}{2}
\frac{\rmd}{\rmd t}\frac{\partial\phi}{\partial x}(t;x,u)&=
\frac{\partial f}{\partial x}(\phi(t;x,u),u)\frac{\partial\phi}{\partial x}(t;x,u),
\quad&
\frac{\partial\phi}{\partial x}(0;x,u)&=I;
\label{eq:eq4dx}\\
\frac{\rmd}{\rmd t}\frac{\partial\phi}{\partial u}(t;x,u)&=
\frac{\partial f}{\partial x}(\phi(t;x,u),u)\frac{\partial\phi}{\partial u}(t;x,u)
+\frac{\partial f}{\partial u}(\phi(t;x,u),u),
\quad&
\frac{\partial\phi}{\partial u}(0;x,u)&=O.
\label{eq:eq4du}
\end{alignat}
Application of a numerical technique simultaneously to 
\eqref{eq:eq4phi}--\eqref{eq:eq4du} gives the values of
$(\partial\phi/\partial x)(jh/M;x,u)$ and $(\partial\phi/\partial u)(jh/M;x,u)$
for $j=1, 2, \ldots, M$.
It is also possible to evaluate the second-order derivatives in a similar way.
In later sections, we need
$\phi(t;x,u)$, $(\partial\phi/\partial x)(t;x,u)$,
$(\partial\phi/\partial u)(t;x,u)$,
$(\partial^2\phi_i/\partial u\partial x)(t;x,u)$, and
$(\partial^2\phi_i/\partial u^2)(t;x,u)$
for $i=1, 2, \ldots, n$.
This means that the total dimension of the differential equations
to be considered is $n+n^2+nm+n^2m+nm^2$.
In Section~\ref{sec:shooting},
we additionally need $(\partial^2\phi_i/\partial x^2)(t;x,u)$
for $i=1, 2, \ldots, n$,
which increases the dimension by $n^3$.

For the original continuous-time plant \eqref{eq:original},
let us write its linearization as
\[
\dot{x}(t)=Ax(t)+Bu(t)
\]
for $A=(\partial f/\partial x)(0,0)$ and $B=(\partial f/\partial u)(0,0)$.
Similarly, the linearization of the discretized plant \eqref{eq:discretized} is
\[
x[k+1]=A_h x[k]+B_h u[k]
\]
for $A_t=(\partial\phi/\partial x)(t;0,0)$ and
$B_t=(\partial\phi/\partial u)(t;0,0)$.
The following relationships follow immediately.

\begin{prop}
\label{prop:linear}
There hold $A_t=\rme^{At}$ and $B_t=\int_0^t \rme^{A\tau}\,\rmd\tau B$.
\end{prop}

\noindent
{\em Proof.}
Substitute $x=0$ and $u=0$ into \eqref{eq:eq4phi} and note $f(0,0)=0$
to see $\phi(t;0,0)=0$ for any $t\geq 0$.
Substitute $x=0$ and $u=0$ into \eqref{eq:eq4dx}
and use $\phi(t;0,0)=0$ and $A=(\partial f/\partial x)(0,0)$.
Then, $(\partial\phi/\partial x)(t;0,0)=\rme^{At}$ follows.
A similar reasoning with \eqref{eq:eq4du} gives 
$(\partial\phi/\partial u)(t;0,0)=\int_0^t\rme^{A\tau}\,\rmd\tau B$.
\qed\bigskip

We make the following assumption.

\begin{assum}
\label{assum:linear}
The original plant \eqref{eq:original} has the linearization such that
$(A,B)$ is stabilizable and $(Q,A)$ is detectable.
\fin
\end{assum}

\noindent
This leads to the following proposition.
Henceforth, we choose the sampling period $h$ so as to have
the property in the proposition.

\begin{prop}
For a small enough sampling period $h$,
the discretized plant \eqref{eq:discretized} has the linearization such that
$(A_h,B_h)$ is stabilizable and $(Q,A_h)$ is detectable.
Then the discrete-time Riccati equation
\begin{equation}	
A_h^\rmT SA_h-S+hQ-A_h^\rmT SB_h(B_h^\rmT SB_h+hR)^{-1}
B_h^\rmT SA_h=O
\label{eq:Riccati}
\end{equation}
has a unique stabilizing solution $S$, which is positive semidefinite.
\end{prop}

\noindent
{\em Proof.}
See \cite[Theorem 3.2.1]{ChF95} for the former statement.
See \cite[Theorem 6.3.2]{ChF95} or \cite[Section 4.1]{Ber00}
for the latter statement.
\qed

\section{Hamiltonian System}
\label{sec:Hamilton}
In order to solve our optimal control problem,
we consider the corresponding Hamiltonian and the Hamiltonian system.

The Hamiltonian \cite{Ber00} for our problem is
\[
H(x[k],u[k],p[k+1])
=\phi(h;x[k],u[k])^\rmT p[k+1]+hx[k]^\rmT Qx[k]+hu[k]^\rmT Ru[k]
\]
with the costate $p[k]\in\Rset^n$.
As a domain of the costate $p[k]$,
we fix some bounded open set $P\subset\Rset^n$ containing the origin.
To simplify the notation,
let us write $x^+[k]=x[k+1]$ and $p^+[k]=p[k+1]$ and
drop the dependence on $k$.
The following assumption is made.

\begin{assum}
\label{assum:convexity}
The Hamiltonian $H(x,u,p^+)$ is convex in $u\in U$
for any $(x,p^+)\in X\times P$.
\fin
\end{assum}

\noindent
This assumption is satisfied for a small enough sampling period $h$
in the following special case.

\begin{prop}
\label{prop:convexity}
Suppose that the original plant \eqref{eq:original} is affine in the input $u(t)$.
Then Assumption~\ref{assum:convexity} is satisfied
for a small enough sampling period $h$.
\end{prop}

\noindent
{\em Proof.}
The Hessian of the Hamiltonian $H(x,u,p^+)$ with respect to $u$ is
\begin{equation}
\frac{\partial^2}{\partial u^2}\phi(h;x,u)^\rmT p^+ +2hR.
\label{eq:convexitypf1}
\end{equation}
Here, the first term is the Hessian of the scalar-valued function
$\phi(h;x,u)^\rmT p^+$.
Note that the Hessian $(\partial^2\phi_i/\partial u^2)(0;x,u)$ is
equal to zero for each of $i=1, 2, \ldots, n$.
Hence, as $h\rightarrow 0$,
\begin{equation}
\frac{1}{h}\frac{\partial^2\phi_i}{\partial u^2}(h;x,u)
\rightarrow
\frac{\rmd}{\rmd t}\frac{\partial^2\phi_i}{\partial u^2}(0;x,u).
\label{eq:convexitypf2}
\end{equation}
The last quantity is in fact equal to $(\partial^2 f_i/\partial u^2)(x,u)$
for each $i=1, 2, \ldots, n$.
This can be seen by
differentiating the first equation of \eqref{eq:eq4du} with respect to $u$
and noting that the derivatives
$(\partial\phi/\partial u)(0;x,u)$ and $(\partial^2\phi_i/\partial u^2)(0;x,u)$
are all equal to zero.
Moreover, the affinity of $f(x,u)$ implies that
the right-hand side quantity in \eqref{eq:convexitypf2}
equals to zero.
The convergence is uniform on $X\times P$ due to the smoothness of $\phi(t;x,u)$.
Hence, the Hessian \eqref{eq:convexitypf1} is positive definite on $X\times P$
for a small enough $h$.
\qed
\bigskip

Henceforth, we write $\phi(h;x,u)$ as $\phi_h(x,u)$ in short.
The minimum principle for our problem is the following \cite{Ber00}:
\begin{align}
x^+&=\frac{\partial H}{\partial p^+}(x,u,p^+)^\rmT=
\phi_h(x,u),
\label{eq:PMPx}\\
0&=\frac{\partial H}{\partial u}(x,u,p^+)^\rmT=
\frac{\partial\phi_h}{\partial u}(x,u)^\rmT p^+ +2hRu,
\label{eq:PMPu}\\
p&=\frac{\partial H}{\partial x}(x,u,p^+)^\rmT=
\frac{\partial\phi_h}{\partial x}(x,u)^\rmT p^+ +2hQx.
\label{eq:PMPp}
\end{align}

\noindent
The equations \eqref{eq:PMPx}--\eqref{eq:PMPp} are
regarded to define some dynamics on the state $(x,p)$
around the origin.
Indeed, we have the following property.

\begin{prop}
\label{prop:Hamilton}
The equations \eqref{eq:PMPx}--\eqref{eq:PMPp}
uniquely determine a smooth function from $(x,p)\in X\times P$
to $(x^+,u,p^+)\in X\times U\times P$ around the origin.
Conversely, 
the equations \eqref{eq:PMPx}--\eqref{eq:PMPp}
uniquely determine a smooth function from $(x^+,p^+)\in X\times P$
to $(x,u,p)\in X\times U\times P$ around the origin.
\end{prop}

\noindent
{\em Proof.}
When $x, p, u, x^+, p^+$ are all equal to zero,
the equations \eqref{eq:PMPx}--\eqref{eq:PMPp} are obviously satisfied.
Suppose first that $(x,p)\in X\times P$ is given and
look for $(x^+,u,p^+)$ satisfying \eqref{eq:PMPx}--\eqref{eq:PMPp}.
Differentiating the right-hand sides of \eqref{eq:PMPu} and \eqref{eq:PMPp}
with respect to $(u,p^+)$ to have
\[
\begin{pmatrix}
\frac{\partial^2}{\partial u^2}\phi_h(x,u)^\rmT p^+ +2hR &
\frac{\partial\phi_h}{\partial u}(x,u)^\rmT \\
\frac{\partial^2}{\partial x\partial u}\phi_h(x,u)^\rmT p^+ &
\frac{\partial\phi_h}{\partial x}(x,u)^\rmT
\end{pmatrix}.
\]
Again, $(\partial^2/\partial u^2)\phi_h(x,u)^\rmT p^+$
is the second derivative of the scalar-valued function
$\phi_h(x,u)^\rmT p^+$.
When $x, p, u, x^+, p^+$ are all equal to zero,
this matrix is reduced to
\[
\begin{pmatrix}
2hR & B_h^\rmT \\ O & A_h^\rmT
\end{pmatrix},
\]
which is nonsingular because $A_h=\rme^{Ah}$ due to Proposition~\ref{prop:linear}.
Hence, the implicit function theorem implies the existence of
a unique mapping from $(x,p)$ to $(u,p^+)$ around the origin
and then to $x^+$ by \eqref{eq:PMPx}.

The converse case where $(x^+,p^+)\in X\times P$ is given
can be similarly considered.
\qed\bigskip

\section{Proposed Method}
\label{sec:method}
Navasca \cite{Nav02,Nav03} considered optimal control of a discrete-time nonlinear
system and showed in particular the existence of a stable
manifold for the associated Hamiltonian system and
its connection to the optimal control.
Application of these results to the present framework gives the following.
Sketch of the proof is presented in the appendix for completeness.

\begin{prop}
\label{prop:manifold}
The Hamiltonian system \eqref{eq:PMPx}--\eqref{eq:PMPp}
has a stable manifold around the origin that can be
expressed as $p=s(x)$.
Here, $s(x)$ is a smooth function satisfying $s(0)=0$ and
$(\partial s/\partial x)(0)=2S$
for the stabilizing solution $S$ of the Riccati equation \eqref{eq:Riccati}.
\end{prop}

\begin{prop}
\label{prop:Lagrange}
For the function $s(x)$ in the previous proposition,
there exists a smooth scalar-valued function $V(x)$ around the origin
such that $(\partial V/\partial x)(x)^\rmT=s(x)$.
\end{prop}

\begin{prop}
\label{prop:HJB}
Let $s(x)$ and $V(x)$ be the functions in the preceding two propositions.
Define the function $u^*(x)$ by the value of $u$ corresponding to
$(x,p=s(x))$ in the Hamiltonian system \eqref{eq:PMPx}--\eqref{eq:PMPp}.
Then, $V(x)$ and $u^*(x)$ satisfy the following equation around the origin:
\[
V(x)
=\min_{u\in U}\Bigl[V(\phi_h(x,u))+hx^\rmT Qx+hu^\rmT Ru\Bigr]
=V(\phi_h(x,u^*(x)))+hx^\rmT Qx+hu^*(x)^\rmT Ru^*(x).
\]
This is the Hamilton--Jacobi--Bellman equation of dynamic programming
and tells that
$V(x)$ is the optimal cost-to-go function for the state $x$
and $u^*(x)$ gives the optimal control.
\end{prop}

Based on the last proposition,
we propose a method to solve our problem.
That is,
we obtain a sufficient number of points $(x,p)$ as well as
the corresponding $u$ on the stable manifold
and approximate the relationship between $x$ and $u$ to have
the function $u^*(x)$.
In particular, we choose $x[N]$ sufficiently close to the origin
for some positive number $N$ and consider the point $(x[N],p[N])$
for $p[N]=2Sx[N]$.
Since this point can be regarded to be on the stable manifold,
so are the points $(x[k],p[k])$, $k=N-1, N-2, \ldots, 0$,
generated by the Hamiltonian system \eqref{eq:PMPx}--\eqref{eq:PMPp}
backward in time.
Thus we can collect desired pairs $(x[k],u[k])$.
Repeat this procedure for various values of $x[N]$
and interpolate the obtained $u$ as a function of $x$.

There are some issues for execution of this procedure.

First, we need to consider how the initial value $x[N]$ is to be chosen.
As will be discussed in the next section,
it is desired to be chosen so that 
the corresponding $x[0]$ matches some prespecified value.
This is accomplished by some shooting method, to be proposed there.

Next, backward simulation of the Hamiltonian system
\eqref{eq:PMPx}--\eqref{eq:PMPp} is discussed.
For a given $(x^+,p^+)$, the nonlinear equations
\eqref{eq:PMPx} and \eqref{eq:PMPu} have to be solved
to have $(x,u)$.
Newton's method is used for this purpose.
\bigskip

\begin{algo}
\label{algo:simulation}\mbox{}\\
Input: the pair $(x[N],p[N])$.\\
Output: a sequence of triplets $(x[k],u[k],p[k])$ for $k=N-1, \ldots, 1, 0$.\\
Procedure:\vspace*{-0.15cm}
\begin{enumerate}[1.]
\item
Set $k:=N-1$.

\item
With $(x^+,p^+)=(x[k+1],p[k+1])$ compute $(x,u,p)$ satisfying
\eqref{eq:PMPx}--\eqref{eq:PMPp} by the following procedure.
\begin{enumerate}[(1)]
\item
Set the initial value $(x^{(0)},u^{(0)})$ in some way.
One possibility is to set $x^{(0)}$ to $x^+=x[k+1]$ and
$u^{(0)}$ to $u[k+1]$.
When $k=N-1$, since $u[k+1]$ is not available,
set $u^{(0)}=-(B_h^\rmT SB_h+hR)^{-1}B_h^\rmT SA_h x^+$
(linear optimal control) instead, for example.

\item
For $j=0, 1, \ldots$, repeat the following until convergence:
\[
\begin{pmatrix} x^{(j+1)} \\ u^{(j+1)} \end{pmatrix}:=
\begin{pmatrix} x^{(j)} \\ u^{(j)} \end{pmatrix}-
\begin{pmatrix}
\frac{\partial\phi_h}{\partial x}(x^{(j)},u^{(j)}) &
\frac{\partial\phi_h}{\partial u}(x^{(j)},u^{(j)}) \\
\frac{\partial^2}{\partial u\partial x}\phi_h(x^{(j)},u^{(j)})^\rmT p^+ &
\frac{\partial^2}{\partial u^2}\phi_h(x^{(j)},u^{(j)})^\rmT p^+ +2hR
\end{pmatrix}^{-1}
\begin{pmatrix}
-x^+ +\phi_h(x^{(j)},u^{(j)}) \\
\frac{\partial\phi_h}{\partial u}(x^{(j)},u^{(j)})^\rmT p^+ +2hRu^{(j)}
\end{pmatrix}.
\]
The necessary derivatives of $\phi_h(x,u)$ can be computed as in Section~\ref{sec:prel}.

\item
Substitute the obtained $(x,u)$ to \eqref{eq:PMPp} to have $p$.
\end{enumerate}

\item
Define $(x[k],u[k],p[k])$ by the obtained $(x,u,p)$.

\item
If $k=0$, output the obtained sequence and stop.
Otherwise, decrease $k$ by one and go back to Step~2.
\fin
\end{enumerate}
\end{algo}

Finally, interpolation of the obtained pairs $(x,u)$ is considered.
When the number of the pairs is medium-size and
the dimensions of $x$ and $u$ are small,
any method such as least-squares approximation with a polynomial
can be used.
When the problem size is large, a more sophisticated method
such as moving least squares and a radial basis function
is preferable \cite{Wen05}.

\section{Shooting Method}
\label{sec:shooting}
In this section, we present a shooting method to choose the value of $x[N]$
so that the corresponding $x[0]$ matches the prespecified value.
As will be seen in Section~\ref{sec:example},
a practical plant can have some important domain of the state
through which the plant moves under control.
Suppose that we choose a sufficient number of points over the domain
and let each of them be $x[0]$ to compute the corresponding optimal input.
Then, through interpolation,
we can obtain the optimal control law $u^*(x)$ for any $x$ in the domain.
This should be more efficient than computing $u^*(x)$
blindly in a large domain.
The idea of the shooting method has been presented in \cite{OiS17}
for optimal control of a continuous-time system.

To present the method,
assume that we have an original trajectory of the Hamiltonian system
\eqref{eq:PMPx}--\eqref{eq:PMPp}
$(x[k],u[k],p[k])$, $k=0, 1, \ldots, N-1$, and $(x[N],p[N])$
on the stable manifold.
Here, $x[N]$ is assumed to be close enough to the origin and
$x[0]$ be not very distant from the prespecified value $x^*$.
We iteratively update the trajectory so that $x[0]$ approaches
the desired value $x^*$.
To this aim, we consider to add a small perturbation $\Delta x[N]$
to $x[N]$.
The perturbation to $p[N]$ is chosen as $\Delta p[N]=2S\Delta x[N]$
so that the point remains on the stable manifold.
We compute the corresponding perturbation that arises
at each point $(x[k],p[k])$ in the trajectory.

Suppose that we have the perturbation $(\Delta x[k+1],\Delta p[k+1])$
for the $(k+1)$st point $(x[k+1],p[k+1])$ and
want to compute the perturbation for the next point
$(x[k],p[k])$, where $k=N-1, \ldots, 1, 0$.
Linearization of the Hamiltonian system \eqref{eq:PMPx}--\eqref{eq:PMPp}
gives
\begin{align}
\Delta x[k+1]&=
\frac{\partial\phi_h}{\partial x}\Delta x[k]+
\frac{\partial\phi_h}{\partial u}\Delta u[k],
\label{eq:perturbx}\\
0&=
\frac{\partial^2}{\partial u\partial x}\phi_h^\rmT p^+\Delta x[k]+
\Bigl(\frac{\partial^2}{\partial u^2}\phi_h^\rmT p^+ +2hR\Bigr)\Delta u[k]+
\Bigl(\frac{\partial\phi_h}{\partial u}\Bigr)^\rmT\Delta p[k+1],
\label{eq:perturbu}\\
\Delta p[k]&=
\Bigl(\frac{\partial^2}{\partial x^2}\phi_h^\rmT p^+ +2hQ\Bigr)\Delta x[k]+
\frac{\partial^2}{\partial x\partial u}\phi_h^\rmT p^+\Delta u[k]+
\Bigl(\frac{\partial\phi_h}{\partial x}\Bigr)^\rmT\Delta p[k+1],
\label{eq:perturbp}
\end{align}
where the derivatives are all evaluated at $(x,u,p^+)=(x[k],u[k],p[k+1])$.
Delete $\Delta u$ and solve them for $\Delta x[k]$ and $\Delta p[k]$.
The result can be expressed as
\[
\begin{pmatrix} \Delta x[k] \\ \Delta p[k] \end{pmatrix}
=
H_k
\begin{pmatrix} \Delta x[k+1] \\ \Delta p[k+1] \end{pmatrix}
\]
with the matrix $H_k$ defined as
\begin{equation}
H_k=
\begin{pmatrix}
I & O \\
\frac{\partial^2}{\partial x^2}\phi_h^\rmT p^+ +2hQ &
\frac{\partial^2}{\partial x\partial u}\phi_h^\rmT p^+
\end{pmatrix}
\begin{pmatrix}
\frac{\partial\phi_h}{\partial x} & \frac{\partial\phi_h}{\partial u} \\
\frac{\partial^2}{\partial u\partial x}\phi_h^\rmT p^+ &
\frac{\partial^2}{\partial u^2}\phi_h^\rmT p^+ +2hR
\end{pmatrix}^{-1}
\begin{pmatrix} I & O \\ O & -\bigl(\frac{\partial\phi_h}{\partial u}\bigr)^\rmT
\end{pmatrix}
+
\begin{pmatrix} O & O \\ O & \bigl(\frac{\partial\phi_h}{\partial x}\bigr)^\rmT
\end{pmatrix}.
\label{eq:Hk}
\end{equation}
The matrix $H_k$ contains the new derivative
$(\partial^2/\partial x^2)\phi_h(x,u)^\rmT p^+$,
which is however computable as in Section~\ref{sec:prel}.

Successive multiplication of the matrix $H_k$ gives
the perturbation for $(x[0],p[0])$, that is,
\[
\begin{pmatrix} \Delta x[0] \\ \Delta p[0] \end{pmatrix}
=
H_0 H_1 \cdots H_{N-1}
\begin{pmatrix} I \\ 2S \end{pmatrix}
\Delta x[N].
\]
Since we want to make $x[0]+\Delta x[0]$ equal to $x^*$,
an appropriate choice of $\Delta x[N]$ should be
\[
\Delta x[N]=\biggl[
(I\ \ O)H_0 H_1 \cdots H_{N-1}
\begin{pmatrix} I \\ 2S \end{pmatrix}\biggr]^{-1}
(x^*-x[0]).
\]
Modify the present $x[N]$ and $p[N]$ by adding
this $\Delta x[N]$ and $2S\Delta x[N]$, respectively,
and update the trajectory with Algorithm~\ref{algo:simulation}
applied to this new initial point.
If the resulting $x[0]$ is close enough to the desired value $x^*$,
we are done.
If this is not the case,
we again consider to add a perturbation to $x[N]$ and
update the trajectory in a similar way.
Since this is basically Newton's method,
convergence can be expected if the original $x[0]$
is not very distant from the desired $x^*$.

This is the idea of the shooting method.
To make it more practical, some numerical techniques are applied.

Since the matrix $H_k$ originates from the Hamiltonian system,
it can have both stable and unstable eigenvalues.
This means that it may have a large condition number and
its successive multiplication may produce a large numerical error.
In order to attenuate the error,
the (thin) QR-factorization
\cite[Section 5.2]{GoV96} \cite[Section 2.9]{SuM03}
should be useful.
First, factorize $(I\ \ 2S)^\rmT$ into
\begin{equation}
\begin{pmatrix} Y_N \\ Z_N \end{pmatrix}R_N
=
\begin{pmatrix} I \\ 2S \end{pmatrix}
\label{eq:QR}
\end{equation}
so that the columns of $(Y_N^\rmT\ \ Z_N^\rmT)^\rmT$ are orthonormal and
the matrix $R_N$ is square and upper-triangular.
Multiply $H_{N-1}$ only to $(Y_N^\rmT\ \ Z_N^\rmT)^\rmT$ and factorize
the resulting matrix as
\[
\begin{pmatrix} Y_{N-1} \\ Z_{N-1} \end{pmatrix}R_{N-1}
=
H_{N-1}\begin{pmatrix} Y_N \\ Z_N \end{pmatrix}
\]
again with the QR-factorization.
Since each column of $(Y_N^\rmT\ \ Z_N^\rmT)^\rmT$ is orthonormal,
the multiplication of $H_{N-1}$ can be made in a more numerically stable manner
than its direct multiplication to $(I\ \ 2S)^\rmT$.
Due to the final form
\[
\begin{pmatrix} \Delta x[0] \\ \Delta p[0] \end{pmatrix}
=
\begin{pmatrix} Y_0 \\ Z_0 \end{pmatrix}
R_0 R_1 \cdots R_N
\Delta x[N],
\]
$\Delta x[N]$ can be computed by
$R_N^{-1}\cdots R_1^{-1}R_0^{-1}Y_0^{-1}(x^*-x[0])$.
Since each $R_k$ is upper-triangular, the multiplication of its inverse can be
computed by simple sequential substitution.

It is possible that the added perturbation $\Delta x[N]$ is too large
and the resulting $x[0]$ does not improve its distance to $x^*$.
In that case, the damping technique can be useful.
That is, we replace the perturbation by its half $(1/2)\Delta x[N]$ and
update the trajectory with Algorithm~\ref{algo:simulation}.
If the resulting $x[0]$ improves the distance to $x^*$, we accept it.
If this is not the case, we again halve the perturbation to $(1/2^2)\Delta x[N]$
and repeat the procedure.

Combination of these ideas gives the following algorithm.

\begin{algo}
\label{algo:shooting}\mbox{}\\[-0.8cm]
\begin{description}
\item[\rm Input:]
a sequence $(x[k],u[k],p[k])$, $k=0, 1, \ldots, N-1$,
and $(x[N],p[N])$ on the stable manifold
such that $x[N]$ is close enough to the origin and
$x(0)$ is not too distant from the desired value $x^*$.
\vspace*{-0.15cm}
\item[\rm Output:]
a sequence $(x[k],u[k],p[k])$, $k=0, 1, \ldots, N-1$,
and $(x[N],p[N])$ on the stable manifold
such that $x[0]$ is close enough to $x^*$.
\end{description}\vspace*{-0.15cm}
Procedure:\vspace*{-0.15cm}
\begin{enumerate}[1.]
\item
Evaluate and store the present distance $d=\|x^*-x[0]\|$
in the Euclid norm.

\item
Apply the QR-decomposition to $(I\ \ 2S)^\rmT$
to have $Y_N$, $Z_N$, and $R_N$ as in \eqref{eq:QR}.

\item
For $k=N-1, \ldots, 1, 0$,
compute $H_k$ through \eqref{eq:Hk}
and apply the QR-decomposition to the product of $H_k$ and
$(Y_{k+1}^\rmT\ \ Z_{k+1}^\rmT)^\rmT$ as
\[
\begin{pmatrix} Y_k \\ Z_k \end{pmatrix}R_k
=
H_k\begin{pmatrix} Y_{k+1} \\ Z_{k+1} \end{pmatrix}.
\]

\item
Determine the perturbation $\Delta x[N]$ by
$R_N^{-1}\cdots R_1^{-1}R_0^{-1}Y_0^{-1}(x^*-x[0])$
and $\Delta p[N]$ by $2S\Delta x[N]$.
With $(x[N]+\Delta x[N],p[N]+\Delta p[N])$ being the new initial point,
produce a trajectory using Algorithm~\ref{algo:simulation}.
If the resulting $x[0]$ improves the distance to $x^*$
as $\|x^*-x[0]\|<d$, accept the produced trajectory
and proceed to the next step.
If this is not the case, replace $\Delta x[N]$ and $\Delta p[N]$
by their halves $(1/2)\Delta x[N]$ and $(1/2)\Delta p[N]$, respectively,
and produce a trajectory again.

\item
If the distance $\|x^*-x[0]\|$ is small enough,
output the present trajectory and stop.
Otherwise, go back to Step~1.
\fin
\end{enumerate}
\end{algo}

\section{Intersample Behavior}
\label{sec:inter}
So far we have considered optimal control of a nonlinear system \eqref{eq:original}
with respect to the objective function \eqref{eq:objective},
which is evaluated only at the sampling instants.
Such objective function can be problematic, however, because
the controlled system may behave badly between the sampling instants.
For sampled-data control of a linear system,
there is a well-established methodology to take account of
the intersample behavior (See \cite{ChF95} and the references therein).
The objective of this section is to extend the method developed so far
in this direction.

Let us replace the objective function \eqref{eq:objective} by
\begin{equation}
\sum_{k=0}^\infty \Bigl[
\int_0^h \phi(t;x[k],u[k])^\rmT Q\phi(t;x[k],u[k])\,\rmd t+hu[k]^\rmT Ru[k]\Bigr].
\label{eq:objectiveinter}
\end{equation}
Since this objective function reflects the system behavior
not only at the sampling instants but also between them,
the resulting control law is expected to give more acceptable control performance.
We will see what needs to be modified in the method
for adaptation to the new objective function.
Let us write the integral in \eqref{eq:objectiveinter}
as $q(x[k],u[k])$ for simplicity.

First we need to modify the Riccati equation \eqref{eq:Riccati},
which gives the optimal control law around the origin.
Due to Proposition~\ref{prop:linear},
the function $\phi(t;x,u)$ can be approximated by
$A_t x+B_t u$ for small $x$ and $u$ with
$A_t=(\partial\phi/\partial x)(t;0,0)=\rme^{At}$ and
$B_t=(\partial\phi/\partial u)(t;0,0)=\int_0^t\rme^{A\tau}\,\rmd\tau B$.
Hence, the function $q(x,u)$ is approximated by
\[
(x^\rmT\ \ u^\rmT)\int_0^h
\begin{pmatrix} A_t^\rmT \\ B_t^\rmT \end{pmatrix}
Q(A_t\ \ B_t)\,\rmd t
\begin{pmatrix} x \\ u \end{pmatrix}.
\]
Let us combine this with the input penalty $hu^\rmT Ru$ and write
\[
\begin{pmatrix} \ol{Q} & \ol{W} \\ \ol{W}^\rmT & \ol{R} \end{pmatrix}
=
\int_0^h \begin{pmatrix} A_t^\rmT \\ B_t^\rmT \end{pmatrix}
Q(A_t\ \ B_t)\,\rmd t
+\begin{pmatrix} O & O \\ O & hR \end{pmatrix}.
\]
This is the weight matrix on the state and the input
in the approximate version of our control problem.
Its value can be computed by numerical integration,
which will be discussed soon.
Now, with this matrix, the Riccati equation \eqref{eq:Riccati}
is modified as
\begin{equation}	
A_h^\rmT\ol{S}A_h-\ol{S}+\ol{Q}-(A_h^\rmT \ol{S}B_h+\ol{W})
(B_h^\rmT \ol{S}B_h+\ol{R})^{-1}
(B_h^\rmT \ol{S}A_h+\ol{W}^\rmT)=O.
\label{eq:Riccati2}
\end{equation}
We assume the existence of its stabilizing solution $\ol{S}$
and use it in place of $S$.

With the new objective function, the Hamiltonian changes to
\[
\ol{H}(x,u,p^+)=\phi_h(x,u)^\rmT p^+ +q(x,u)+hu^\rmT Ru.
\]
The Hamiltonian system accordingly changes to
\begin{align}
x^+&=\frac{\partial\ol{H}}{\partial p^+}(x,u,p^+)^\rmT
=\phi_h(x,u),
\label{eq:PMPx2}\\
0&=\frac{\partial\ol{H}}{\partial u}(x,u,p^+)^\rmT
=\frac{\partial\phi_h}{\partial u}(x,u)^\rmT p^+
+\frac{\partial q}{\partial u}(x,u)^\rmT+2hRu,
\label{eq:PMPu2}\\
p&=\frac{\partial\ol{H}}{\partial x}(x,u,p^+)^\rmT
=\frac{\partial\phi_h}{\partial x}(x,u)^\rmT p^+
+\frac{\partial q}{\partial x}(x,u)^\rmT.
\label{eq:PMPp2}
\end{align}
These equations determine a unique dynamics in the space of $(x,p)$
around the origin for a small enough sampling period $h$.
Indeed, the right-hand side of \eqref{eq:PMPu2} and \eqref{eq:PMPp2}
differentiated by $(u,p^+)$ is
\[
\begin{pmatrix} 2\ol{R} & B_h^\rmT \\ 2\ol{W} & A_h^\rmT \end{pmatrix}
\]
for $x,p,u,x^+,p^+$ all equal to zero.
This matrix is invertible if and only if so are $2\ol{R}$ and
$A_h^\rmT-2\ol{W}(2\ol{R})^{-1}B_h^\rmT$.
Because the latter is true for a small enough $h$,
the same reasoning as in the proof of Proposition~\ref{prop:Hamilton}
implies that
the mapping from $(x,p)$ to $(u,p^+)$ is uniquely determined
and thus so is the mapping to $x^+$.
The situation is similar on the mapping from $(x^+,p^+)$ to $(x,u,p)$.

Propositions~\ref{prop:manifold}--\ref{prop:HJB} hold as well
after appropriate changes.
Hence it should be possible to obtain the optimal control law.
In order to have a trajectory on the stable manifold,
the initial point $(x[N],p[N])$ should be chosen by
$p[N]=2\ol{S}x[N]$ with the stabilizing solution $\ol{S}$ of
the modified Riccati equation \eqref{eq:Riccati2}.
Algorithm~\ref{algo:simulation} needs to be modified so as to be consistent
with the new Hamiltonian system \eqref{eq:PMPx2}--\eqref{eq:PMPp2}.
In particular, the linear optimal control used in Step~2~(1)
is replaced by
$u^{(0)}=
-(B_h^\rmT\ol{S}B_h+\ol{R})^{-1}(B_h^\rmT\ol{S}A_h+\ol{W}^\rmT)x^+$.
Newton's method in Step~2~(2) should be replaced by
\[
\begin{pmatrix} x^{(i+1)} \\ u^{(i+1)} \end{pmatrix}:=
\begin{pmatrix} x^{(i)} \\ u^{(i)} \end{pmatrix}-
\begin{pmatrix}
\frac{\partial\phi_h}{\partial x} &
\frac{\partial\phi_h}{\partial u} \\
\frac{\partial^2}{\partial u\partial x}\phi_h^\rmT p^+
+\frac{\partial^2 q}{\partial u\partial x} &
\frac{\partial^2}{\partial u^2}\phi_h^\rmT p^+ 
+\frac{\partial^2 q}{\partial u^2}+2hR
\end{pmatrix}^{-1}
\begin{pmatrix}
-x^+ +\phi_h(x^{(i)},u^{(i)}) \\
\bigl(\frac{\partial\phi_h}{\partial u}\bigr)^\rmT p^+
+\bigl(\frac{\partial q}{\partial u}\bigr)^\rmT+2hRu^{(i)}
\end{pmatrix},
\]
where the derivatives are all evaluated at $(x^{(i)},u^{(i)})$.
Here we need evaluation of the derivatives of
$q(x,u)=\int_0^h \phi(t;x,u)^\rmT Q\phi(t;x,u)\,\rmd t$,
which is possible with numerical integration.
For example,
evaluation of the derivative $(\partial q/\partial u)(x,u)$ can be
made by the composite Simpson's rule \cite[Section 7.5]{SuM03}:
\begin{align*}
\frac{\partial q}{\partial u}(x,u)
&=\int_0^h 2\phi(t;x,u)^\rmT Q\frac{\partial\phi}{\partial u}(t;x,u)\,\rmd t\\
&\approx
\frac{h}{3M}\biggl[
2\phi(0;x,u)^\rmT Q\frac{\partial\phi}{\partial u}(0;x,u)
+4\sum_{j=1}^{M/2-1}\phi\bigl(\frac{2jh}{M};x,u\bigr)^\rmT Q
\frac{\partial\phi}{\partial u}\bigl(\frac{2jh}{M};x,u\bigr)\\
&\quad+8\sum_{j=1}^{M/2}\phi\bigl(\frac{(2j-1)h}{M};x,u\bigr)^\rmT Q
\frac{\partial\phi}{\partial u}\bigl(\frac{(2j-1)h}{M};x,u\bigr)
+2\phi(h;x,u)^\rmT Q\frac{\partial\phi}{\partial u}(h;x,u)
\biggr]
\end{align*}
for an even positive number $M$.
It requires the values of the derivative $(\partial\phi/\partial u)(t;x,u)$ at $t=ih/M$,
$i=1, 2, \ldots, M$,
which can be computed by the Runge--Kutta method as in Section~\ref{sec:prel}.

\begin{rem}
\label{rem:Simpson}
The composite Simpson's rule has the numerical error in the order of $\rmO(M^{-4})$
and is suitable to use with the classical fourth-order Runge--Kutta method,
whose numerical error is also in the order of $\rmO(M^{-4})$.
The composite trapezoidal rule, which is more common for numerical integration,
produces a larger numerical error
$\rmO(M^{-2})$ and is not suitable.
\fin
\end{rem}

\noindent
In Step~3 of Algorithm~\ref{algo:simulation},
the computed $(x,u)$ should be substituted into \eqref{eq:PMPp2}
in place of \eqref{eq:PMPp} to give the value of $p$.
Here, we need $(\partial q/\partial x)(x,u)$,
which again is computable with numerical integration.

The shooting method can be used, too, with the new objective function.
To this end, evaluation of $(\partial^2 q/\partial x^2)(x,u)$ is necessary,
which is possible similarly to that of other derivatives.

\begin{figure}
\begin{center}
\setlength{\unitlength}{1cm}
\begin{picture}(0,5.5)(0,0.5)
\put(0,0){\makebox(0,0)[b]{\includegraphics[height=5.5cm]{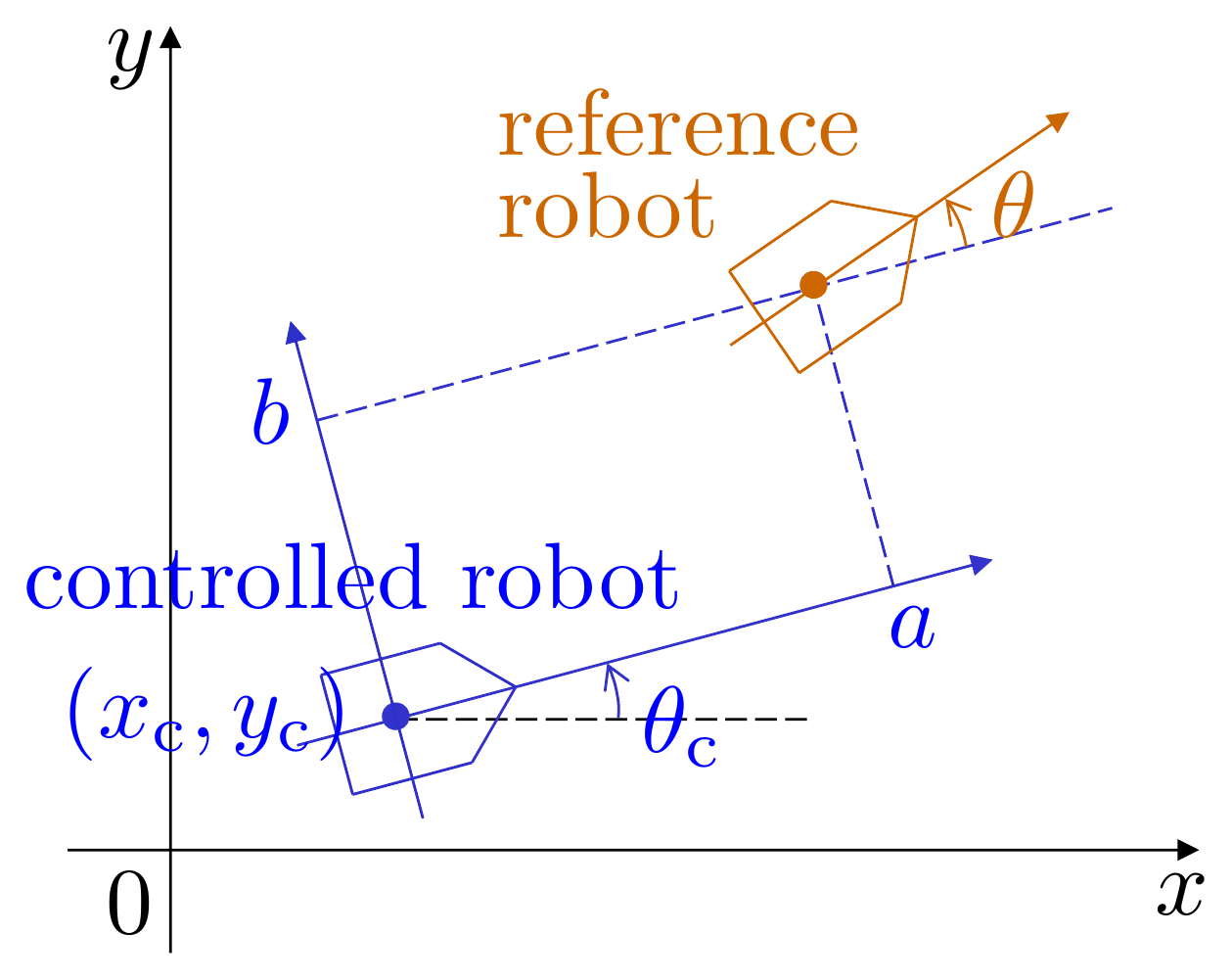}}}
\end{picture}
\end{center}
\caption{Schematic view of the tracking control of a mobile robot.
The objective is to choose the linear and angular velocities
of the controlled robot so that its position $(x_\rmc,y_\rmc)$ and
direction $\theta_\rmc$ become equal to those of the reference robot.
The position and the direction of the reference robot
relative to the controlled robot are $(a,b)$ and $\theta$, respectively.}
\label{fig:mobilerobot}
\end{figure}

\section{Example}
\label{sec:example}
For illustration of the proposed method,
tracking control of a wheeled mobile robot is considered
in the framework of \cite{KKMN90}.
The position of the controlled robot is $(x_\rmc,y_\rmc)$ in the $xy$-coordinates
and its direction is $\theta_\rmc$ measured counterclockwise
from the positive direction of the $x$-axis (Figure~\ref{fig:mobilerobot}).
The linear velocity $v_\rmc$ and the angular velocity $\omega_\rmc$
can be specified independently.
The objective of control is to make the position $(x_\rmc,y_\rmc)$
and the direction $\theta_\rmc$ of the controlled robot
equal to those of the reference robot $(x_\rmr,y_\rmr,\theta_\rmr)$,
which is driven by the known linear and angular velocities $v_\rmr$
and $\omega_\rmr$.
The state of the dynamics is chosen as
\[
\begin{pmatrix} a\\ b\\ \theta \end{pmatrix}=
\begin{pmatrix}
\cos\theta_\rmc & \sin\theta_\rmc & 0 \\
-\sin\theta_\rmc & \cos\theta_\rmc & 0 \\
0 & 0 & 1 \end{pmatrix}
\begin{pmatrix}
x_\rmr-x_\rmc \\ y_\rmr-y_\rmc \\ \theta_\rmr-\theta_\rmc
\end{pmatrix},
\]
where $(a,b)$ stands for the relative position of the reference robot
in the coordinate system fixed to the controlled robot
and $\theta$ is the direction of the reference robot
relative to that of the controlled robot.
With the relative velocities
$v=v_\rmr-v_\rmc$ and $\omega=\omega_\rmr-\omega_\rmc$ for inputs,
one can write the dynamics as
\[
\frac{\rmd}{\rmd t}
\begin{pmatrix} a(t)\\ b(t)\\ \theta(t) \end{pmatrix}=
\begin{pmatrix}
v_\rmr(t)(\cos\theta(t)-1)+b(t)\omega_\rmr(t)+v(t)-b(t)\omega(t) \\
v_\rmr(t)\sin\theta(t)-a(t)\omega_\rmr(t)+a(t)\omega(t) \\
\omega(t)
\end{pmatrix}.
\]
Suppose that the reference robot has $(x_\rmr(t),y_\rmr(t),\theta_\rmr(t))=(0,0,0)$
at the initial time $t=0$ and moves in the positive direction of the
$x$-axis with the constant speed $v_\rmr(t)\equiv 1\,[\rmm/\rms]$ and
$\omega_\rmr(t)\equiv 0\,[\text{rad}/\rms]$.
On the other hand, the controlled robot has
$(x_\rmc(t),y_\rmc(t),\theta_\rmc(t))=(0,0,-\pi)$ at $t=0$,
that is, it is located at the same position as the reference robot
but in the opposite direction.
Hence the initial state is $(a(0)\ \ b(0)\ \ \theta(0))^\rmT=(0\ \ 0\ \ \pi)^\rmT$.
We want to make it converge to the origin by appropriate choice of
the inputs $v(t)$ and $\omega(t)$.
A sampled-data control law is designed for the purpose
with the sampling period $h=1\,[\rms]$ and the weight matrices
$Q$ and $R$ being identities.
Note that, if the inputs $v(t)$ and $\omega(t)$ are generated
by the sampled-data control and thus are piecewise constant,
the inputs $v_\rmc(t)$ and $\omega_\rmc(t)$ actually given to the
controlled robot also become piecewise constant. 

\begin{figure}
\begin{center}
\setlength{\unitlength}{1cm}
\begin{picture}(0,7)(0,0.2)
\put(0,0){\makebox(0,0)[b]{\includegraphics[height=7cm]{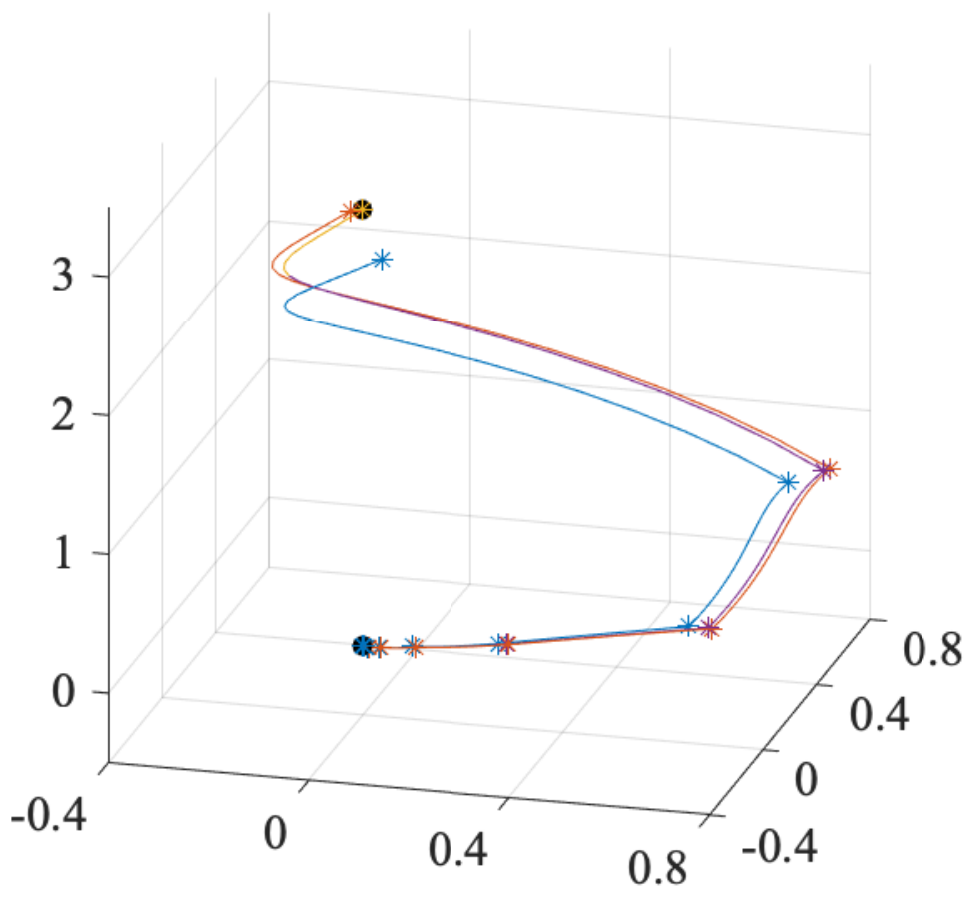}}}
\put(-1.1,0.1){\makebox(0,0){$a\,[\rmm]$}}
\put(3.4,0.9){\makebox(0,0){$b\,[\rmm]$}}
\put(-3.9,3.2){\makebox(0,0){$\theta\,[\text{rad}]$}}
\put(-0.4,5){\makebox(0,0){\color{blue}1st}}
\put(-1.5,5.5){\makebox(0,0){\color{red}2nd}}
\put(-0.55,5.5){\makebox(0,0){\color{brown}3rd}}
\put(0.05,5.5){\makebox(0,0){\color{purple}4th}}
\put(-1.4,2){\makebox(0,0)
{$\displaystyle\begin{pmatrix}0\\[-0.15cm]0\\[-0.15cm]0\end{pmatrix}$}}
\put(-1,6.3){\makebox(0,0)
{$\displaystyle\begin{pmatrix}0\\[-0.15cm]0\\[-0.15cm]\pi\end{pmatrix}$}}
\end{picture}
\end{center}
\caption{The trajectory updates made by the shooting method.
The original trajectory is updated three times and
gives a trajectory reaching the desired point $(0\ \ 0\ \ \pi)^\rmT$.
Here, the third and the fourth trajectories are too close
to distinguish in this scale.
The intersample behavior is presented for reference
though not required by the method.}
\label{fig:robotshoot}
\end{figure}
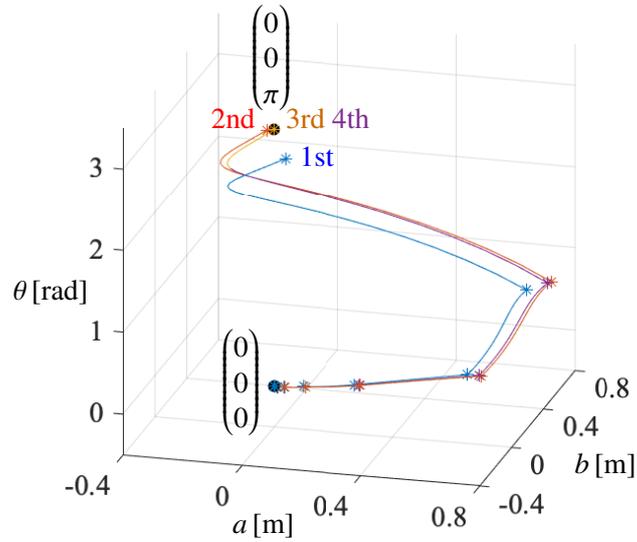

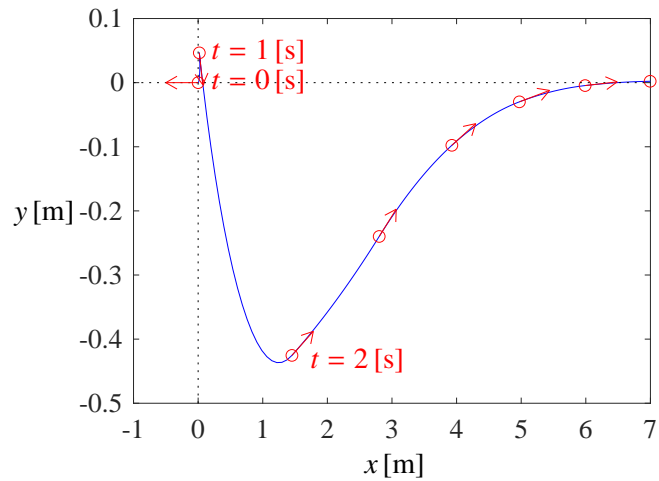
\begin{figure}
\begin{center}
\setlength{\unitlength}{1cm}
\begin{picture}(0,6)(0,0.2)
\put(0,0){\makebox(0,0)[b]{\includegraphics[height=6cm]{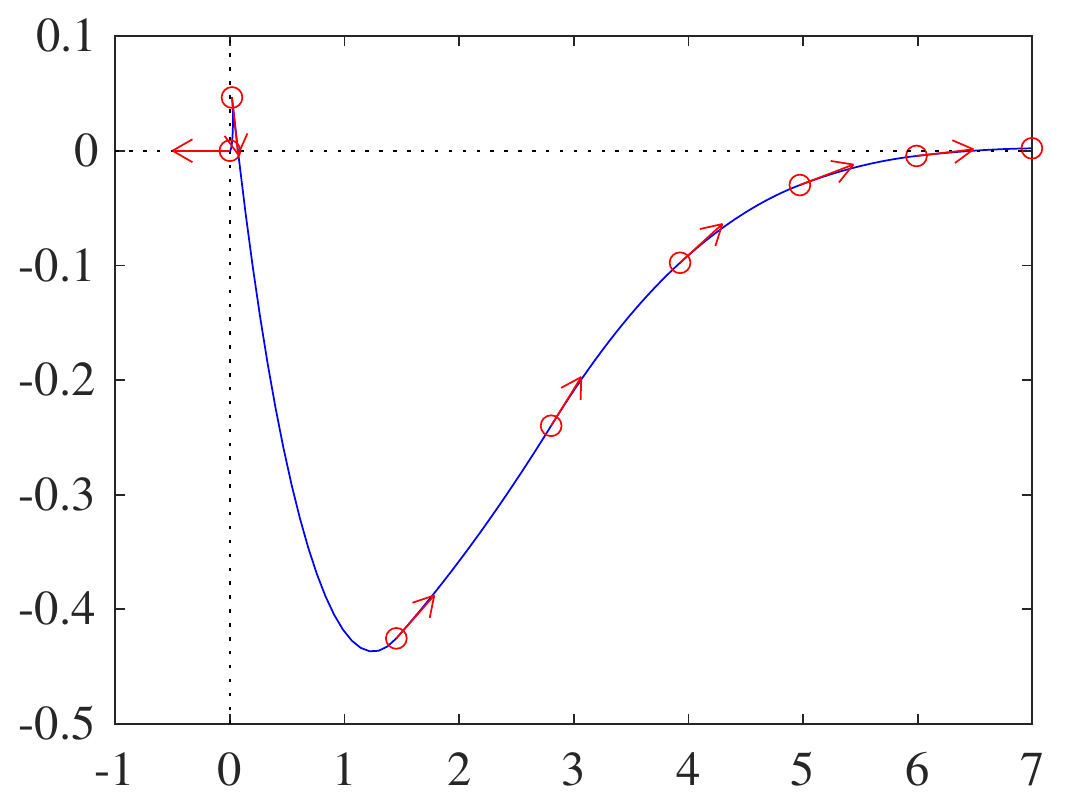}}}
\put(0.3,-0.2){\makebox(0,0){$x\,[\rmm]$}}
\put(-4.3,3.15){\makebox(0,0){$y\,[\rmm]$}}
\put(-2.1,4.9){\makebox(0,0)[l]{\color{red}$t=0\,[\rms]$}}
\put(-2.1,5.3){\makebox(0,0)[l]{\color{red}$t=1\,[\rms]$}}
\put(-0.8,1.2){\makebox(0,0)[l]{\color{red}$t=2\,[\rms]$}}
\end{picture}
\end{center}
\caption{The computed trajectory of the controlled robot
in the original coordinate $(x_\rmc,y_\rmc)$.
The position and the direction at the sampling instants are
shown by the circles and the arrows, respectively.
The robot successfully catches up the reference robot
with a long sampling period $h=1\,[\rms]$.}
\label{fig:robotpos}
\end{figure}

\begin{figure}
\begin{center}
\setlength{\unitlength}{1cm}
\begin{picture}(0,6)(0,0.2)
\put(0,0){\makebox(0,0)[b]{\includegraphics[height=6cm]{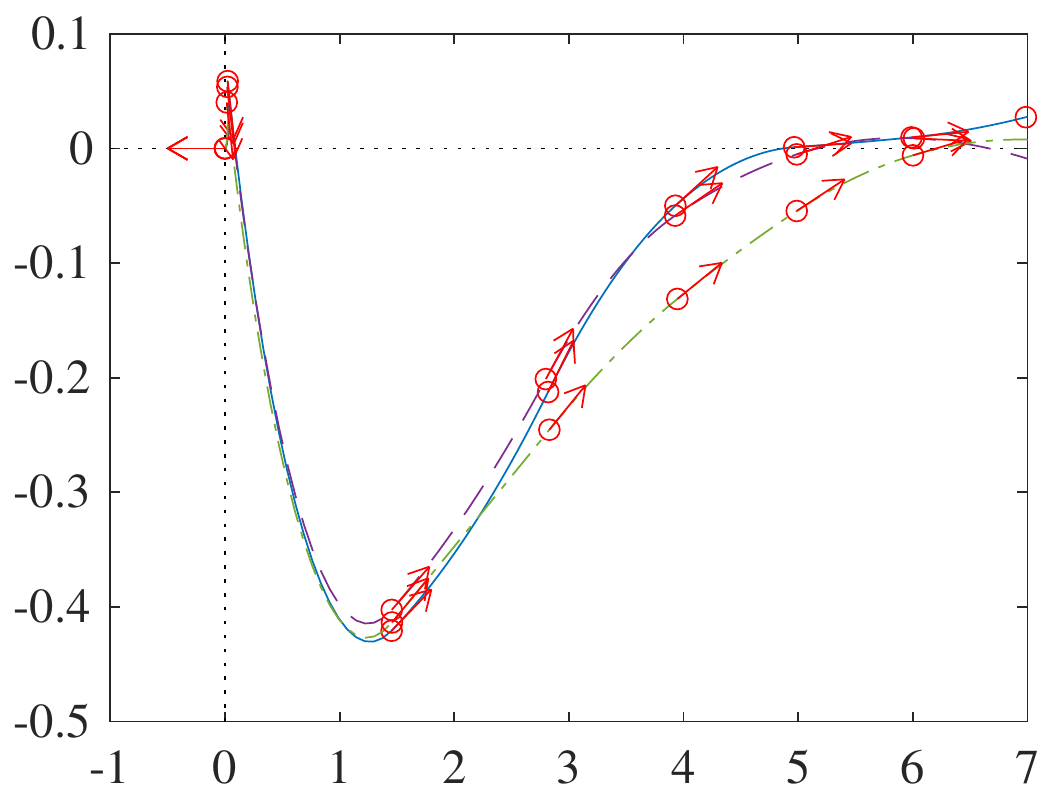}}}
\put(0.3,-0.2){\makebox(0,0){$x\,[\rmm]$}}
\put(-4.3,3.15){\makebox(0,0){$y\,[\rmm]$}}
\end{picture}
\end{center}
\caption{Tracking control under the measurement noise.
Three realizations are presented by the lines of different color and style.
Tracking is successfully made though some fluctuation is found.}
\label{fig:robotposnoise}
\end{figure}

\begin{figure}
\begin{center}
\setlength{\unitlength}{1cm}
\begin{picture}(0,6)(0,0.2)
\put(0,0){\makebox(0,0)[b]{\includegraphics[height=6cm]{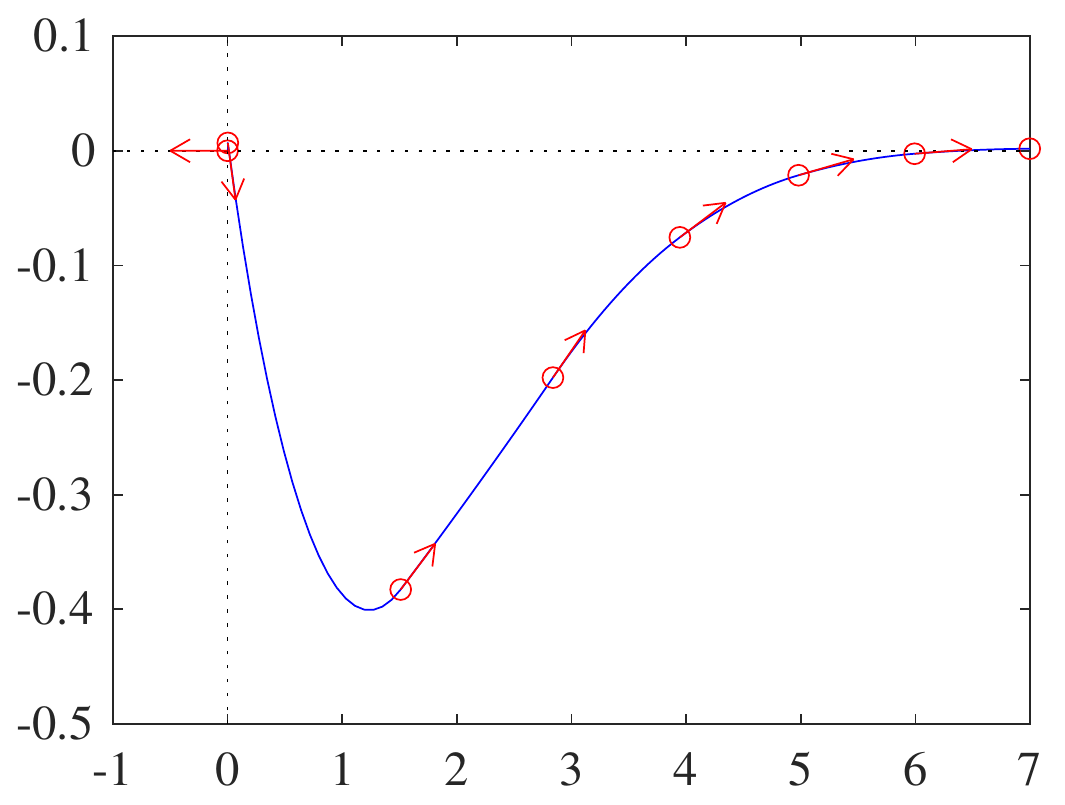}}}
\put(0.3,-0.2){\makebox(0,0){$x\,[\rmm]$}}
\put(-4.3,3.15){\makebox(0,0){$y\,[\rmm]$}}
\put(-2.1,4.8){\makebox(0,0)[l]{\color{red}$t=0\,[\rms]$}}
\put(-2.1,5.15){\makebox(0,0)[l]{\color{red}$t=1\,[\rms]$}}
\put(-0.8,1.5){\makebox(0,0)[l]{\color{red}$t=2\,[\rms]$}}
\end{picture}
\end{center}
\caption{The trajectory of the controlled robot
with the intersample behavior considered.
Deviation from the $x$-axis is smaller than
that of the trajectory in Figure~\ref{fig:robotpos}.}
\label{fig:robotposinter}
\end{figure}

As described in Section~\ref{sec:method},
by choosing some $x[N]$ close to the origin and
applying Algorithm~\ref{algo:simulation},
we can have a trajectory on the stable manifold.
Because our initial state is
$(a(0)\ \ b(0)\ \ \theta(0))^\rmT=(0\ \ 0\ \ \pi)^\rmT$,
we want a trajectory whose $x[0]$ coincides with this vector.
For the purpose, we invoke the shooting method of Algorithm~\ref{algo:shooting}
with the desired state $x^*=(0\ \ 0\ \ \pi)^\rmT$.
The result is shown in Figure~\ref{fig:robotshoot}.
Here, the intersample behavior of the state is presented for reference.
After three updates, the desirable trajectory that reaches
$(0\ \ 0\ \ \pi)^\rmT$ was obtained.
In the figure, the last update is too small to recognize.
Figure~\ref{fig:robotpos} shows the obtained trajectory
in the original coordinate $(x_\rmc,y_\rmc)$,
where the position at the sampling instants is expressed by the circles
and the direction by the arrows.
Note that the scale of the $y$-coordinate is presented $10$ times larger.
We can see that the robot first goes backward with changing the direction
counterclockwise and then goes forward in the negative direction of the $y$-axis.
It gradually changes the direction and finally catches up the reference robot.
Although the inputs are kept constant between the sampling instants,
still the control objective is achieved successfully.

In order to have a control law in a feedback form,
we applied the shooting method with setting $x^*$ to various values
and obtained the corresponding input $u$.
In particular, we added Gaussian noise of standard deviation $0.2$
to each component of $x[0]$, $x[1]$, $x[2]$, $x[3]$ of the finally obtained
trajectory in Figure~\ref{fig:robotshoot}
and also to the origin and produced $100$ values of $x^*$.
With $x^*$ set to each value of them, the corresponding $u$ was obtained.
The obtained values of $u$ were approximated by a third-order polynomial
of $x$, which is our control law.
The obtained control law was applied to tracking control.
In order to see its sensitivity to the noise,
we added Gaussian noise of standard deviation $0.02$ to
each component of the measured state and then passed it
to the control law.
The result is presented in Figure~\ref{fig:robotposnoise}.
Even with the measurement noise, the tracking was successfully made
though some fluctuation is found.

Figure~\ref{fig:robotposinter} shows the optimal trajectory of the robot
when the intersample behavior is taken into account.
Compared with the trajectory of Figure~\ref{fig:robotpos},
the deviation from the $x$-axis is smaller both
in the positive and negative directions of the $y$ axis.
Indeed, in the objective function \eqref{eq:objectiveinter},
the sum of the first term, {\em i.e.}, the state penalty,
is $5.52$ in this trajectory, which is smaller than
the value $5.86$ in the trajectory of Figure~\ref{fig:robotpos}.
On the other hand, the sum of the second term of the objective function,
{\em i.e.}, the input penalty,
is larger in the present trajectory
and thus the overall value of the objective function is
not very different.
In particular, it is $13.90$ in the present trajectory while
$13.93$ in the trajectory of Figure~\ref{fig:robotpos}.

\section{Conclusion}
\label{sec:concl}
In this paper, optimal sampled-data control is considered
for a nonlinear system with the stable-manifold approach.
The approach can be adapted for sampled-data control
thanks to the works of Navasca and
can be carried out with extensive use of numerical techniques.
A shooting method is proposed for systematic choice of an initial point of
a trajectory.
The extension is considered for the intersample behavior of the system
to be taken into account.
An example shows the efficacy of the proposed method.

\section*{Appendix}
Outline of the proof is presented
for Propositions~\ref{prop:manifold}--\ref{prop:HJB}.
See \cite{Nav02,Nav03} for details.
\bigskip

{\em (Outline of the Proof of Proposition~\ref{prop:manifold})}
The existence of the stable manifold is shown by the contraction mapping theorem.
Expansion in $x$ gives $s(0)=0$ and $(\partial s/\partial x)(0)=2S$.
\qed
\bigskip

{\em (Outline of the Proof of Proposition~\ref{prop:Lagrange})}
Due to the form of the Hamiltonian system \eqref{eq:PMPx}--\eqref{eq:PMPp},
the symplectic form satisfies
$\sum_{i=1}^n \rmd p_i\wedge\rmd x_i=\sum_{i=1}^n \rmd p^+_i\wedge\rmd x^+_i$.
This means $\sum_{i=1}^n \rmd p_i\wedge\rmd x_i=0$ on the stable manifold
because $(x[k],p[k])\rightarrow 0$ as $k\rightarrow\infty$ there.
Noting that $p$ is a function of $x$ on the stable manifold, substitute
$\rmd p_i=\sum_{j=1}^n(\partial p_i/\partial x_j)\rmd x_j$
to have $(\partial p_i/\partial x_j)-(\partial p_j/\partial x_i)=0$
for any $i$ and $j$.
Poincar\'{e}'s lemma then implies the existence of the desired $V(x)$.
\qed
\bigskip

{\em (Outline of the Proof of Proposition~\ref{prop:HJB})}
Substitute $p=s(x)=(\partial V/\partial x)(x)^\rmT$ and $u=u^*(x)$
into \eqref{eq:PMPu} to have
\begin{align*}
0
&=\frac{\partial\phi_h}{\partial u}(x,u^*(x))^\rmT
\frac{\partial V}{\partial x}(\phi_h(x,u^*(x))^\rmT+2hRu^*(x)\\
&=\frac{\partial}{\partial u}\bigl[V(\phi_h(x,u))+hx^\rmT Qx+hu^\rmT Ru\bigr]^\rmT
\bigr|_{u=u^*(x)}.
\end{align*}
This shows that $u=u^*(x)$ minimizes the quantity in the bracket.

Similarly, substitution of $p=s(x)=(\partial V/\partial x)(x)^\rmT$ and $u=u^*(x)$
into \eqref{eq:PMPp} gives
\begin{align*}
\frac{\partial V}{\partial x}(x)^\rmT
&=\frac{\partial\phi_h}{\partial x}(x,u^*(x))^\rmT
\frac{\partial V}{\partial x}(\phi_h(x,u^*(x))^\rmT+2hQx\\
&=\frac{\partial}{\partial x}\bigl[V(\phi_h(x,u))+hx^\rmT Qx+hu^\rmT Ru\bigr]^\rmT
\bigr|_{u=u^*(x)}\\
&=\frac{\partial}{\partial x}\bigl[V(\phi_h(x,u^*(x)))+hx^\rmT Qx+hu^*(x)^\rmT Ru^*(x)\bigr]^\rmT.
\end{align*}
See \cite[Lemma 3.3.1]{Ber00} for the derivation of the last expression.
This shows that $V(x)-V(\phi_h(x,u^*(x)))-hx^\rmT Qx-hu^*(x)^\rmT Ru^*(x)$
is a constant independent of $x$.
This constant is actually zero because $u^*(x)=0$ for $x=0$.
\qed

\end{document}